\documentclass[letterpaper, 10 pt, conference]{ieeeconf}

\IEEEoverridecommandlockouts                             
\usepackage{amsthm}
\usepackage{algorithm}
\usepackage[noend]{algpseudocode}
\usepackage{booktabs} 
\usepackage{array}
\usepackage{balance}
\usepackage[noadjust]{cite}
\usepackage{graphicx}
\usepackage{hyperref}
\newtheorem{hypothesis}{Hypothesis}
\newtheorem{exploratory}{Exploratory Question}

\title{\LARGE \bf
Beyond Attention: Investigating the Threshold Where Objective Robot Exclusion Becomes Subjective
}

\author{Clarissa Sabrina Arlinghaus$^{1*}$, Ashita Ashok$^{2}$, Ashim Mandal$^{2}$, Karsten Berns$^{2}$, and Günter W. Maier$^{1}$
\thanks{$^{1}$Bielefeld University, Bielefeld, Germany}%
\thanks{$^{2}$RPTU Kaiserslautern-Landau, Kaiserslautern, Germany}
\thanks{\tt\small *clarissa\_sabrina.arlinghaus@uni-bielefeld.de}%
\thanks{This work has been submitted to the IEEE for possible publication. 
    Copyright may be transferred without notice, after which this version 
    may no longer be accessible.}
}

\begin{document}

\maketitle
\thispagestyle{empty}
\pagestyle{empty}

\begin{abstract}
As robots become increasingly involved in decision-making processes (e.g., personnel selection), concerns about fairness and social inclusion arise. This study examines social exclusion in robot-led group interviews by robot Ameca, exploring the relationship between objective exclusion (robot’s attention allocation), subjective exclusion (perceived exclusion), mood change, and need fulfillment. In a controlled lab study (\textit{N} = 35), higher objective exclusion significantly predicted subjective exclusion. In turn, subjective exclusion negatively impacted mood and need fulfillment but only mediated the relationship between objective exclusion and need fulfillment. A piecewise regression analysis identified a critical threshold at which objective exclusion begins to be perceived as subjective exclusion. Additionally, the standing position was the primary predictor of exclusion, whereas demographic factors (e.g., gender, height) had no significant effect. These findings underscore the need to consider both objective and subjective exclusion in human-robot interactions and have implications for fairness in robot-assisted hiring processes.

\end{abstract}

\section{INTRODUCTION}
With robot's growing presence in modern workplace \cite{otting2022let}, they are now also being integrated into personnel selection processes, where they can participate in majority decision-making to select candidates \cite{masjutin2022we} 
or conduct robot-mediated job interviews \cite{kumazaki2017android,norskov2020applicant}.
While these applications can increase efficiency and consistency, concerns about bias and discrimination remain \cite{chen2023ethics, kochling2020discriminated}.
At the same time, robot-led interviews may offer valuable training opportunities, particularly for migrants and non-native speakers facing linguistic barriers \cite{van2019social}. Initial studies suggest that migrants perceive robot-mediated training experiences positively \cite{ashok2025thanks}.

\begin{figure}[ht]
\centerline{\includegraphics[width=\linewidth]{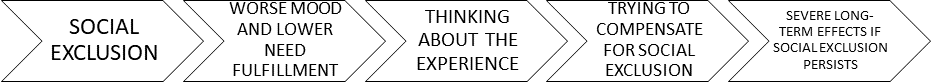}}
\caption{Temporal Need-Threat Model (TNTM) (adapted from \cite{williams2009ostracism})}
\label{fig:TNTM}
\vspace{-3mm}
\end{figure}

In this study, we examine social exclusion in robot-mediated job interviews using a simulated interview setting. Specifically, we investigate how robot attention distribution (objective exclusion) relates to subjective exclusion, mood changes, and need fulfillment. Our work builds on prior research on social exclusion in human-robot interactions (HRIs) \cite{mongile2023if, stachnick2024isolated, rosenthal2023seriously, strassmann2024don}, 
and is theoretically grounded in the Temporal Need-Threat Model (TNTM), which explains how social exclusion negatively affects mood and need fulfillment (see Figure \ref{fig:TNTM})\cite{williams2009ostracism}.
TNTM studied through the Cyberball paradigm, shows that receiving fewer ball passes worsens mood and need fulfillment, even when co-players are robots \cite{williams2006cyberball, hartgerink2015ordinal, erel2021excluded}. Our study extends this framework to conversational settings, where speaking opportunities in group discussions mirror ball-passing dynamics in Cyberball.

Previous research indicates that individuals feel excluded when robots direct fewer questions to them \cite{buttner2024does, mongile2023if} 
or when they are unable to participate in a conversation because the robots communicate in a language, they do not understand \cite{stachnick2024isolated,rosenthal2023seriously}.
We explored robot-led job interviews as a novel setting for studying exclusion, using continuous measures instead of binary inclusion/exclusion conditions. By analyzing natural variations in robot attention, we identified thresholds where objective exclusion becomes subjective. To investigate this transition, we pose the following research question:

\begin{exploratory}
At what level of objective exclusion do individuals begin to feel subjectively excluded?
\end{exploratory}

Attribution is how individuals explain events and experiences \cite{sanders2024attribution} 
and plays a key role in coping with exclusion \cite{williams2009ostracism}. 
A preceding study suggests that robot exclusion, unlike human exclusion, is attributed to technical limitations or programming constraints rather than interpersonal rejection or workplace bullying, potentially moderating its psychological effects \cite{arlinghaus2025asymmetrical}. 
To examine attribution in robot-led interviews, participants selected a new standing position for a second round and explained their choice. We also assessed whether they attribute interaction outcomes to positioning or conversational content, providing insights into whether exclusion is perceived as self-related (internal) or robot-related (external). An open-ended feedback section captured additional explanations of exclusion experiences. Thus, we formulate our second research question:

\begin{exploratory}
How do participants explain their experience of objective exclusion by the robot?
\end{exploratory}

Additionally, we examined whether individual factors (e.g., age, gender, height, standing position) influenced the likelihood of exclusion. Identifying risk factors was crucial for ensuring fair and inclusive robot-assisted selection in group interview settings. By analyzing whether demographic or spatial variables contributed to exclusion, we aimed to inform the development of equitable and unbiased hiring technologies. Thus, we formulate our final research question:

\begin{exploratory}
Are there specific individual factors (e.g., age, gender, height, physical position/angle) that increase the likelihood of being excluded by the robot?
\end{exploratory}
\begin{figure*}[h]
  \centering
  \begin{minipage}{0.41\textwidth}
    \includegraphics[width=\linewidth]{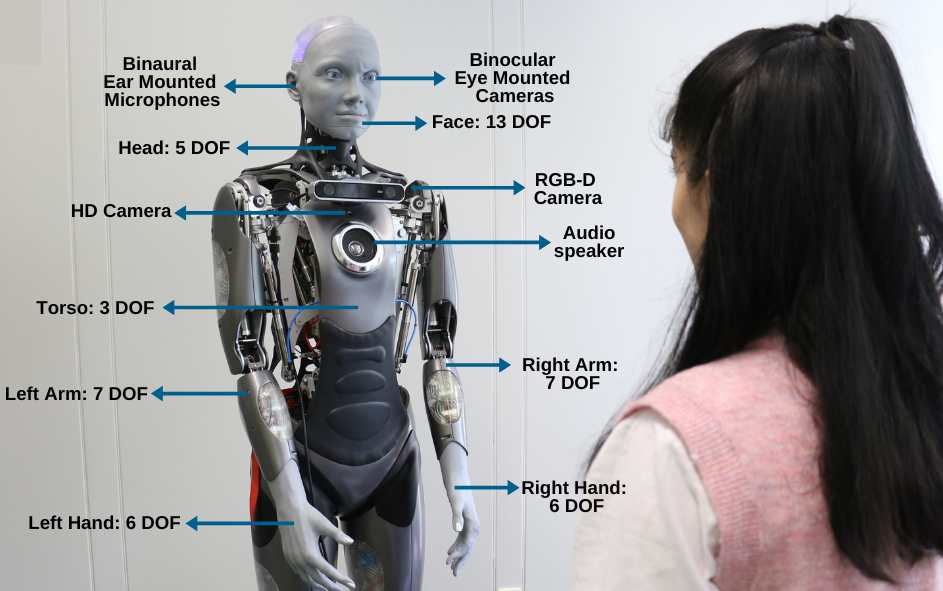}
  \end{minipage}
  \begin{minipage}{0.58\textwidth}
    \includegraphics[width=\linewidth]{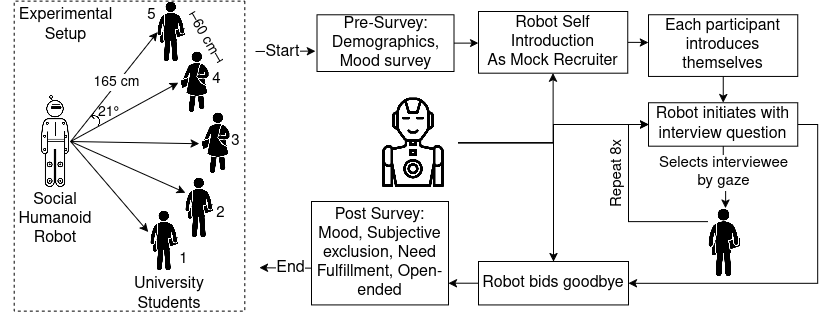}
  \end{minipage}
  \caption{Social humanoid robot Ameca from Engineered Arts (left) used in the exclusion study; mock interview flow with the robot (right).}
  \vspace{-3mm}
\label{fig:combined}
\end{figure*}

Understanding these factors is crucial, as persisting exclusion in social interactions, intentional or unintentional, can have severe psychological consequences \cite{williams2009ostracism}. In the context of robot-mediated job interviews, it is important to examine whether reduced robot attention translates into feelings of exclusion and, in turn, affects the emotional state and fundamental psychological needs of the participants. Building on this, we aim to investigate not only the presence of such exclusion but also its psychological consequences. Specifically, we expect that reduced robot attention (objective exclusion) will lead to an increased sense of being ignored (subjective exclusion), which in turn affects mood and need fulfillment.

Therefore, we hypothesize that in our proposed setting of robot-mediated job interviews with social humanoid robot Ameca\footnote{\url{https://rrlab.cs.rptu.de/en/robots/ameca}}(see Figure \ref{fig:combined}(left)), reduced robot attention (objective exclusion) increases self-reported feelings of being ignored (subjective exclusion), thereby negatively impacting mood and need fulfillment.
\begin{hypothesis}
Higher levels of objective exclusion worsen mood, mediated by subjective exclusion.
\end{hypothesis}
\begin{hypothesis}
Higher objective exclusion reduces need fulfillment, mediated by subjective exclusion.
\end{hypothesis}

Our study advances the understanding of exclusion dynamics in HRIs, specifically in robot-assisted interviews. While prior research focused on subjective exclusion, we systematically compare objective and subjective exclusion to uncover underlying mechanisms. If certain individuals receive less attention, they may be disadvantaged, reinforcing existing inequalities and restricting employment access. Given the link between unemployment and social exclusion \cite{pohlan2019unemployment}
, ensuring bias-free hiring technologies is essential. To mitigate these risks, we advocate for rigorous evaluation of robot-assisted hiring tools to prevent discrimination.

\section{METHODS}

This study was pre-registered on the Open Science Framework\footnote{\url{https://doi.org/10.17605/OSF.IO/HYB2S}}.

\subsection{Sample}
The required sample size was determined via G*Power 3.1 for linear multiple regression (fixed model, \( R^2 \) deviation from zero; \( f² = 0.35 \) (large effect); \( \alpha = 0.05 \); power = 0.80; two predictors), yielding a minimum of 31 participants. In February 2025, we successfully recruited 35  \([N_m=N_f=17, N_o=1)\) international students, primarily from India (\textit{N} = 24, 68.6\%), which is the second largest nationality after natives \cite{ashok2025thanks}. Participants were aged 22–33 years (\textit{M} = 26.43, \textit{SD} = 2.27) and had a height range of 1.30 m to 1.85 m (\textit{M} = 1.65 m, \textit{SD} = 11.46 cm). A majority (54.3\%, \textit{N=19}) were in the final stages of their studies, expecting to graduate in 2025, with German proficiency levels ranging from A1 to C1, with A2 (\textit{N = 12}) and B1 (\textit{N = 11}). All participants provided informed consent and received a 5-euro Best-Choice voucher as compensation for their participation. No participants were excluded from the analyses.

\subsection{Procedure}
This study was conducted at RPTU Kaiserslautern (Germany) in collaboration with Bielefeld University, with ethical approval obtained from both institutions (RPTU: No. 69; Bielefeld: No. 2025-020-S).

Participants were recruited under the premise of a German-language job interview training for non-native speakers. While the interview was in German, questionnaires were provided in English to reduce cognitive load and anxiety. A structured interview format was used, as it enhances reliability and validity \cite{levashina2014structured}. Sessions began with introductions, followed by eight randomized situational and behavioral questions \cite{NorthCentralStateCollege2016}(translated into German), known for their high predictive validity \cite{taylor2002asking}. Each session included five standing participants in a mock group interview, with Ameca autonomously selecting whom to ask the interview question based on Algorithm \ref{algo:eyegaze}. This algorithm outlines the process of initializing eye gaze listeners and dynamically shifting Ameca’s gaze between participants during the interview. Previous studies have found discrepancies in the robot's gaze-tracking algorithm, thereby inspiring this study to look in-depth at possible AI biases. The algorithm continuously updates eye targets, adjusts gaze direction, and incorporates random gaze shifts when no target is detected. Video recordings enabled post-hoc analysis of attention distribution, with objective exclusion quantified by question frequency. Each session lasted approx. 40 minutes, including the interview, questionnaires, and debriefing (see Figure \ref{fig:combined}(right)).

\begin{algorithm}
\caption{Eyegaze Control Algorithm for HR Interview}
\begin{algorithmic}[1]
\State Initialize variables and participant positions (1 to 5)
\Procedure{OnActivate}{}
    \State Set camera brightness, add gaze listeners, introduce robot
\EndProcedure
\Procedure{OnDeactivate}{}
    \State Remove gaze listeners
\EndProcedure
\Procedure{OnLookUpdate}{}
    \State Update/reset eye target
\EndProcedure
\Procedure{OnEyesUpdate}{} (executed every frame)
    \State Set/clear target position
\EndProcedure
\For{each of 8 questions}
    \If{target exists}
        \State Select target from output of OnEyesUpdate(), adjust eye angles
        \State Update gaze, ask question, update target
    \EndIf
    \State Wait for response
\EndFor
\Procedure{RandomGaze}{}
    \State Occasionally shift gaze if no target
\EndProcedure
\State End experiment, deactivate gaze control
\end{algorithmic}
\label{algo:eyegaze}
\end{algorithm}
\vspace{-2mm}

\subsection{Measures}

The following section outlines the study variables, with all instructions adjusted to fit the job interview context.

\begin{itemize}
\item \textbf{Objective Exclusion}: Objective exclusion is assessed by analyzing the robot’s attention during the interview as follows: Objective Exclusion = 1 - (Number of questions and responses directed to a participant / Total number of questions and responses to all participants).
\item \textbf{Subjective Exclusion}: Participants rate two items (e.g., “I was ignored”) on a 5-point Likert scale (1 = Strongly disagree, 5 = Strongly agree) \cite{williams2009ostracism}, assessing their perceived exclusion during the robot interaction.  
\item \textbf{Mood Change}: Participants rate eight items (e.g., "happy", "sad") on a 5-point Likert scale (1 = Not at all, 5 = Extremely) \cite{williams2009ostracism} before and after the interview. Mood Change is operationalized as post-measure minus pre-measure. 
\item \textbf{Need Fulfillment}: Participants rate four items (e.g., “invisible – recognized”) on a 9-point Likert scale \cite{rudert2016s}.
\item \textbf{Standing Position}: Participants select one of five positions before the interview and may choose a new one for a hypothetical second round.
\item \textbf{Open-Ended Feedback}: Participants justify their position choice, assess whether position or responses influenced the robot more, and provide open-ended feedback.
\end{itemize}

\begin{figure*}[h]
  \centering
  \begin{minipage}{0.49\textwidth}
    \includegraphics[width=\linewidth]{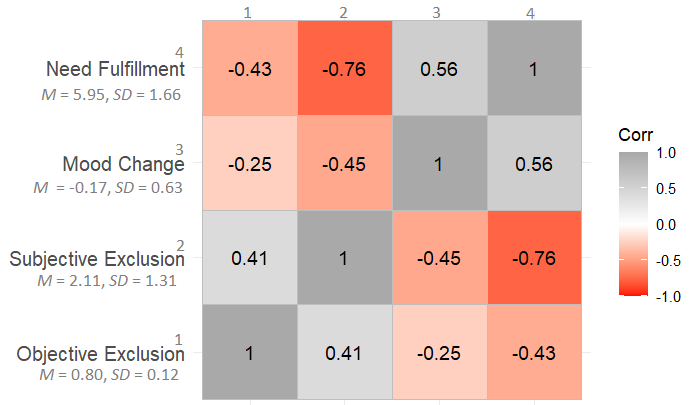}
  \end{minipage}
  \begin{minipage}{0.49\textwidth}
    \includegraphics[width=\linewidth]{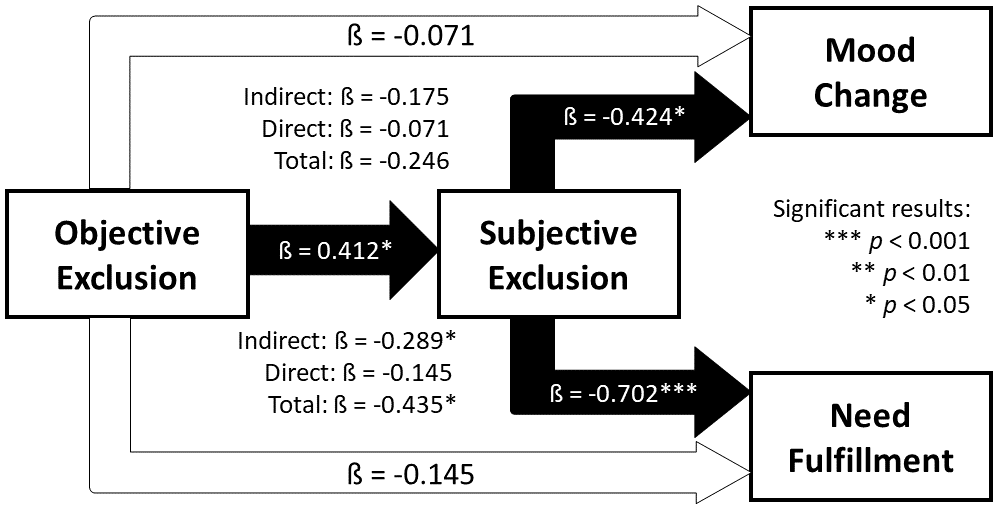}
  \end{minipage}
  \caption{Higher correlation indicating potential mediation effects.}
  \vspace{-3mm}
\label{fig:corr&mediation}
\end{figure*}
\subsection{Analyses}
We assessed normality, linearity, homoscedasticity, and multicollinearity before conducting statistical analyses in R 4.4.3 (R Studio 2024.04.0).
The data set (\textit{N} = 35) contained no extreme outliers {(+-3*IQR)}.

\begin{itemize}
\item \textbf{Mediation Analysis}: Two mediation models tested whether objective exclusion indirectly influenced mood changes and need fulfillment via subjective exclusion, using bootstrapped mediation (5000 resamples, Bollen-Stine test, BCa correction).
\item \textbf{Threshold Analysis}: Piecewise regression identified the inflection point where objective exclusion triggers subjective exclusion.
\item \textbf{Qualitative Analysis}: Open-ended feedback analyzed using LLM-Assisted Inductive Categorization \cite{arlinghaus2024inductive}.
\item \textbf{Risk Factors}: Multiple regressions examined whether age, gender, language proficiency, height, and standing position predicted objective or subjective exclusion.

\end{itemize}

\section{RESULTS}
\subsection{Mediation Analyses}
Robust mediation analyses revealed that objective exclusion significantly predicted subjective exclusion (\ss = 0.412, \textit{SE} = 2.174, \textit{p} = .031). In turn, subjective exclusion significantly predicted both mood change (\ss = -0.424, \textit{SE} = 0.083, \textit{p} = .015) and need fulfillment (\ss = -0.702, \textit{SE} = 0.140, \textit{p} $<$ .001). However, the direct effect of objective exclusion was not significant for either mood change (\ss = -0.071, \textit{SE} = 0.714, \textit{p} = .588) or need fulfillment  (\ss = -0.145, \textit{SE} = 1.676, \textit{p} = .213). Similarly, the indirect effect of objective exclusion on mood change via subjective exclusion was not significant (\ss = -0.175, \textit{SE} = 0.636, \textit{p} = .135). In contrast, the indirect effect on need fulfillment was significant (\ss = -0.289, \textit{SE} = 2.097, \textit{p} = .047). The total effect of objective exclusion was not significant for mood change (\ss = -0.246, \textit{SE} = 0.749, \textit{p} = .074), but reached significance for need fulfillment (\ss = -0.435, \textit{SE} = 2.486, \textit{p} = .012). These results are visually depicted in Figure \ref{fig:corr&mediation}.

\begin{figure}[!b]
\centerline{\includegraphics[width=\linewidth]{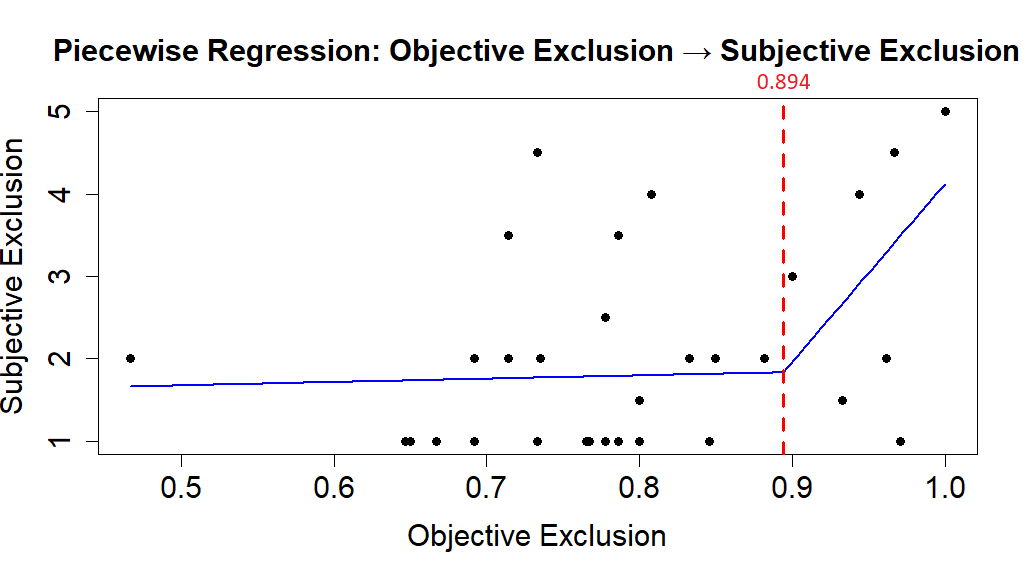}}
\caption{Results of Piecewise Regression.}
\label{fig:piecewise}
\vspace{-3mm}
\end{figure}

\subsection{Threshold Analysis}
A piecewise regression identified a critical threshold at 0.894 for Objective Exclusion, beyond which Subjective Exclusion increased more sharply (see Figure \ref{fig:piecewise}). 

\subsection{Qualitative Analysis}

\begin{figure*}[h]
  \centering
  \begin{minipage}{0.49\textwidth}
    \includegraphics[width=\linewidth]{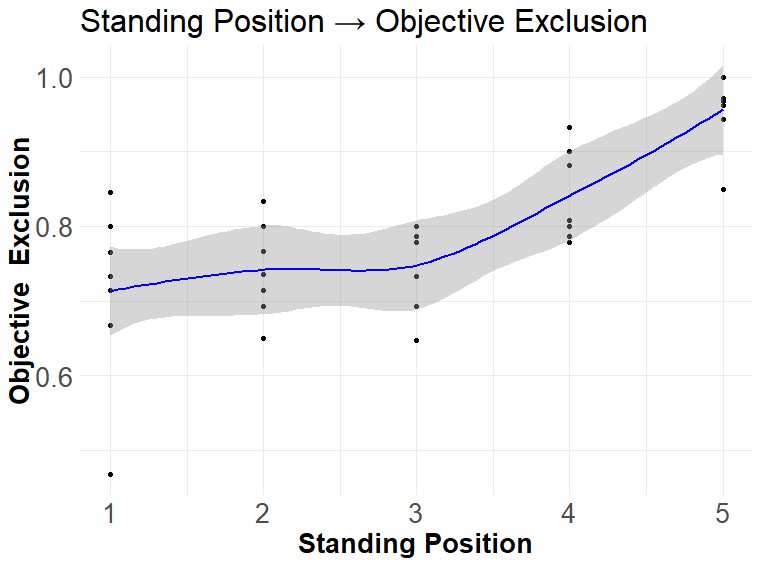}
  \end{minipage}
  \begin{minipage}{0.49\textwidth}
    \includegraphics[width=\linewidth]{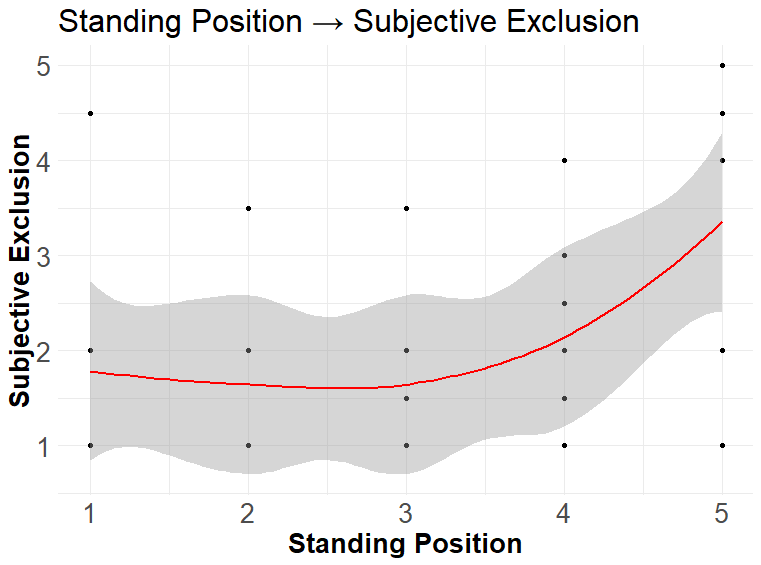}
  \end{minipage}
  \caption{Standing Position as a Risk Factor for Robot-Induced Exclusion.}
\label{fig:standingPos}
\end{figure*}

For a (hypothetical) second interview round, Position 4 was the least selected (\textit{N = 3}), while Position 2 was the most chosen (\textit{N = 10}). Seventeen participants retained their initial position, while 18 switched. Reasons for switching included robot behavior observations (\textit{N = 10}, e.g., “The robot seemed to orient more towards its right and had more eye contact with the person standing in this position”), strategic positioning (\textit{N = 10}, e.g., “Standing directly in front of its gaze might increase my chances of capturing her attention”), personal preference (\textit{N = 5}, e.g., “I would want to be more involved and present in the second and would want to redeem my confidence by giving better responses”), and random selection (\textit{N = 10}, e.g., “There is no specific reason. I am fine with standing anywhere”). Regarding factors influencing HRI, 22 participants identified standing position (e.g., “Standing position is of greater influence”), 7 emphasized response content (e.g., “The content is more important”), and 4 cited both (e.g., “I think both have an equal impact on the flow of conversation”). Two mentioned other factors (e.g., “My height”). In post-experiment feedback, 12 participants commented on question-response dynamics (e.g., “More interactive, where I can also ask some queries back to her”), 9 on gaze/eye contact (e.g., “The eye contact should be held with the person speaking!”), 6 on speech clarity (e.g., “Since I am not very good at German, I would want the robot to speak more clearly and a little more loudly”), 2 provided positive feedback (e.g., “It was good”), and 6 refrained from commenting (e.g., “No”).

\subsection{Risk Factors}

Two multiple linear regression analyses were conducted to examine potential risk factors for exclusion. Standing position was the only significant predictor of both objective exclusion (\ss = 0.057, \textit{SE} = 0.011, \textit{p} $<$ .001) and subjective exclusion (\ss = 0.475, \textit{SE} = 0.159, \textit{p} = .006), indicating that individuals’ location in the group influenced both their actual and perceived exclusion. The model for objective exclusion explained 56.1\% of the variance (\textit{F}(5,27) = 6.91, \textit{p} $<$ .001), whereas the model for subjective exclusion accounted for 27.2\% but was not statistically significant (\textit{F}(5,27) = 2.01, \textit{p} = .109).  A quadratic regression showed that standing position predicted objective exclusion in a curvilinear fashion (\textit{p} = .020, R² = 60.4\%), while the effect for subjective exclusion was not significant (\textit{p} = .062). To visualize the relationship for both outcomes, locally estimated scatterplot smoothing (LOESS) was applied (see Figure \ref{fig:standingPos}).

In contrast, gender, age, height, and German proficiency showed no significant effects (p-values ranging from .626 to .992). An extended model including interaction terms indicated that gender significantly interacted with age (\textit{p} = .004) and standing position (\textit{p} = .019). The interaction between gender and height nearly approached significance (\textit{p} = .054). In contrast, no interaction effects significantly predicted subjective exclusion (p-values ranging from .329 to .888). Finally, the standing position itself was not systematically associated with any individual characteristics (p-values ranging from .147 to .835), indicating that participants did not systematically select their position based on demographics.

\section{DISCUSSION}
This study investigated whether objective exclusion leads to declines in mood and need fulfillment and whether this effect is mediated by subjective exclusion. Additionally, we explored at what level of objective exclusion individuals start to perceive subjective exclusion, how participants explain their experience of exclusion, and whether certain individual characteristics increase the likelihood of being excluded by the robot.

\subsection{The Role of Subjective Exclusion for Negative Effects}
\textbf{Results showed that objectively excluded participants also felt subjectively excluded}. This aligns with previous research, where individuals reported subjective exclusion when they received fewer ball passes \cite{erel2021excluded}, were asked fewer questions \cite{buttner2024does, mongile2023if}, were disadvantaged in argumentation \cite{hitron2022ai}, faced language barriers \cite{stachnick2024isolated, rosenthal2023seriously}, or were ignored in direct requests \cite{strassmann2024don}. However, objective exclusion does not automatically impair mood; rather, the crucial factor is whether individuals perceive themselves as being excluded. Our findings indicate that subjective exclusion mediates the relationship between objective exclusion and need fulfillment but not between objective exclusion and mood change.
While previous studies have shown that objective exclusion affects psychological needs \cite{hartgerink2015ordinal}, our findings highlight subjective exclusion as the driving factor, emphasizing its importance as a core variable and manipulation check in future research.

\subsection{Objective Exclusion Turning into Subjective Exclusion}
\textbf{Study identified a critical threshold for objective exclusion (0.89), beyond which subjective exclusion increased sharply}. This suggests that individuals may tolerate up to 10\% more exclusion before perceiving it as socially meaningful, highlighting a threshold effect that could inform interventions to prevent exclusion in social settings. Future research should consider a threshold when designing exclusion manipulations, investigating different group sizes, and exploring whether a threshold also exists for adverse psychological effects.

\subsection{Explaining the Experience of Exclusion}
\textbf{Results found that the robot's inclusion and exclusion behaviors are predominantly externally attributed to the standing position}. High external attribution has also been reported in previous research. \cite{arlinghaus2025asymmetrical}. Additionally, the current study also found a higher proportion of internal attributions, with some participants believing that their German-language responses influenced the interaction - anxiety commonly experienced by foreign language learners \cite{ashok2025thanks}. This suggests that individual perceptions of robot exclusion may be shaped by contextual or self-inflicted factors, warranting further exploration.

\begin{algorithm}
\caption{Rectified and Fair Eyegaze Control Algorithm}
\begin{algorithmic}[1]
\Procedure{OnEyesUpdate}{} (executed every frame)
    \If{\textit{human face is detected}}
        \If{\textit{eye contact} is detected}
            \State Get gaze coordinates \((x, y)\)
            \State target $\gets$ \Call{SelectTargetFairly}{}
            \State Adjust gaze to target
            \State Ask random question from question list
        \Else
            \State return \textit{"No eye contact"}
        \EndIf
    \Else
        \State return \Call{RandomGaze}{}
    \EndIf
\EndProcedure

\Procedure{SelectTargetFairly}{}
    \State Identify participants with lowest attention count
    \State Break ties randomly
    \State Increment attention count for selected participant
    \State \Return \textit{selected participant}
\EndProcedure
\end{algorithmic}
\label{algo:eyegaze_updated}
\end{algorithm}

\subsection{Risk Factors for Objective and Subjective Exclusion}
\textbf{Regression analyses further supported participants' impression that exclusion was mainly influenced by standing position}. Interviewees in position 5 (see Figure \ref{fig:standingPos}) received less attention and felt more excluded, yet 17\% still chose it for a second interview. Some aimed to reintegrate excluded peers by stepping back themselves, allowing others to receive more attention - a behavior also observed during interviews (e.g., encouraging them to speak) and aligning with prior findings \cite{mongile2023if}.
These results emphasize the impact of spatial positioning in group dynamics, as physical location affected participants’ likelihood of being addressed or ignored. Considering spatial dynamics in robot-led interviews is essential for ensuring fairness in HRI. To support equitable interactions, HRI developers should design adaptive AI systems that account for spatial bias and ensure balanced engagement across all interlocutors. A rectified algorithm is presented in Algorithm \ref{algo:eyegaze_updated} that maintains fair target selection in group interactions such as mock group interviews with robots.
However, gender, age, height, and language proficiency did not significantly predict exclusion, which is encouraging for equitable HRIs. Although prior research warns of gender biases in robot interactions \cite{hitron2022ai,buttner2024does}, our findings suggest that women were not disadvantaged in group conversations. Moreover, women did not systematically choose less favorable standing positions, ruling out self-selection bias.

\section{CONCLUSION}
Overall, our findings contribute to the understanding of exclusion in HRIs, demonstrating that subjective perception is key in translating robotic exclusion into psychological outcomes. Objective exclusion by robot alone does not automatically worsen mood or need fulfillment; rather, individuals exhibit an exclusion threshold, beyond which subjective exclusion significantly intensifies. Our results highlight standing position as a primary determinant of exclusion, while demographic factors play a negligible role. Future research should explore whether robot behavior adaptations could mitigate exclusion, particularly for individuals in less favorable positions, and how spatial positioning interacts with group size and task structure to foster more inclusive interactions. By considering both subjective perceptions and contextual influences, future studies can refine strategies for designing fair and socially inclusive human-robot interactions.

\vspace{-3mm}
\section*{ACKNOWLEDGMENT}

SAIL\footnote{\url{www.sail.nrw}} is funded by the Ministry of Culture and Science of the State of North Rhine-Westphalia under the grant no NW21-05A.

\section*{DATA AVAILABILITY}
The data that support the findings of this study are openly available here: \url{https://osf.io/s7k5u/files/osfstorage}

\bibliographystyle{IEEEtran}

\bibliography{root}

\end{document}